\documentclass[usenatbib]{mn2e}

\usepackage{amssymb,amsmath,aas_macros}
\usepackage{graphicx}
\usepackage{multirow}
\usepackage{natbib}

\def\cite{\citep}

\newcommand{\be}{\begin{equation}}
\newcommand{\ee}{\end{equation}}
\def\bea{\begin{eqnarray}}
\def\ena{\end{eqnarray}}

\usepackage{url}
\usepackage{float}
\usepackage{bm}
\footnotesize
\newdimen\minuswidth    
\setbox0=\hbox{$-$}
\minuswidth=\wd0
\catcode`@=\active
\def@{\kern\minuswidth}
\newdimen\digitwidth    
\setbox0=\hbox{\rm0}
\digitwidth=\wd0
\catcode`!=\active
\def!{\kern\digitwidth}
\normalsize

\usepackage[usenames,dvipsnames]{color}

\footnotesize
\newdimen\digitwidth    
\setbox0=\hbox{\rm0}
\digitwidth=\wd0
\catcode`!=\active
\def!{\kern\digitwidth}
\normalsize
\begin{document}

\title[Gravitational wave memory effect]
{Searching for gravitational wave memory bursts with the Parkes Pulsar Timing Array} \makeatletter
\author[J.B. Wang et al.]{J. B. Wang,$^{1,2,3}$
G. Hobbs,$^{3}$
W. Coles,$^4$
R. M. Shannon,$^{3}$
X. J. Zhu,$^{5,3}$
D. R. Madison,$^6$
\newauthor
M. Kerr,$^3$
V. Ravi,$^{7,3}$
M. J. Keith,$^{8}$
R.N. Manchester,$^{3}$
Y. Levin,$^{9}$
M. Bailes,$^{10}$
\newauthor
N. D. R. Bhat,$^{11}$
S. Burke-Spolaor,$^{12}$
S. Dai,$^{13,3}$
S. Os{\l}owski,$^{14,15}$
W. van Straten,$^{10}$
\newauthor
L. Toomey,$^3$
N. Wang,$^{1,16}$
L. Wen$^{5}$
\\
$^1$Xinjiang Astronomical Observatory, Chinese Academy of Science,
150 Science 1-Street, Urumqi, Xinjiang, China, 830011 \\
$^2$University of Chinese Academy of Sciences, Beijing, China, 100049 \\
$^3$CSIRO Astronomy and Space Science, Australia Telescope National Facility,
PO Box 76, Epping, NSW 1710, Australia \\
$^4$Electrical and Computer Engineering, University of California at San Diego, La Jolla, California, U.S.A. \\
$^{5}$University of Western Australia, 35 Stirling Hwy, Crawley, WA 6009, Australia\\
$^6$Department of Astronomy and Center for Radiophysics and Space Research, Cornell University, Ithaca, NY 14850, USA\\
$^{7}$School of Physics, University of Melbourne, Vic 3010, Australia \\
$^{8}$Jodrell Bank Centre for Astrophysics, School of Physics and Astronomy, The University of Manchester, Manchester M13 9PL, UK\\
$^9$School of Physics, Monash University, P. O. Box 27, Vic 3800, Australia \\
$^{10}$Centre for Astrophysics and Supercomputing, Swinburne University of Technology, P.O. Box 218, Hawthorn, VIC 3122 \\
$^{11}$International Centre for Radio Astronomy Research, Curtin University, Bentley, WA 6102, Australia \\
$^{12}$California Institute of Technology, Pasadena, 1200 E California Blvd, CA 91125, U.S.A.\\
$^{13}$Department of Astronomy, School of Physics, Peking University, Beijing 100871, China\\
$^{14}$Max-Planck-Institut f{\"u}r Radioastronomie, Auf dem H{\"u}gel 69, D-53121 Bonn, Germany\\
$^{15}$Department of Physics, Universit{\"a}t Bielefeld Universit{\"a}tsstr. 25 D-33615 Bielefeld, Germany\\
$^{16}$Key Laboratory of Radio Astronomy, Chinese Academy of Science, Nanjing, China, 210008 \\
 }

\date{printed \today}
\maketitle

\begin{abstract}
\indent Anisotropic bursts of gravitational radiation produced by events such as super-massive black hole mergers leave permanent imprints on space. Such gravitational wave ``memory'' (GWM) signals are, in principle, detectable through pulsar timing as sudden changes in the apparent pulse frequency of a pulsar. If an array of pulsars is monitored as a GWM signal passes over the Earth, the pulsars would simultaneously appear to change pulse frequency by an amount that varies with their sky position in a quadrupolar fashion. Here we describe a search algorithm for such events and apply the algorithm to approximately six years of data from the Parkes Pulsar  Timing Array. We find no GWM events and set an upper bound on the rate for events which could have been detected.  We show, using simple models of black hole coalescence rates, that this non-detection is not unexpected.

\end{abstract}

\begin{keywords}
gravitational waves --- methods: data analysis --- pulsars: general
\end{keywords}

\section{Introduction}

 As supermassive black hole binary (SMBHB) systems coalesce they are expected to produce gravitational wave (GW) emission.   At the time of coalescence a permanent change in the space-time metric will propagate away from the source \citep{Payne1983,Christodoulou1991,Blanchet1992,Thorne1992,Favata2009}.  The permanent change is known as the ``gravitational wave memory" (GWM) effect.  Throughout this paper we mainly consider GWM events caused by SMBHB coalescences.  However, GWM events can also come from other sources such as cosmic strings, supernovae or during the flyby of massive objects (Pshirkov, Baskaran \& Postnov 2010).

The passage of a GWM past the Earth or a pulsar will cause a change in the observed frequency of that pulsar's rotation. By observing a sufficiently large number of stable millisecond pulsars it is expected that an unambiguous detection of the GWM effect could be made.  In this paper we describe a GWM search algorithm and apply it to the recent Parkes Pulsar Timing Array (PPTA) data set (Manchester et al., 2013).

As part of the pulsar-timing technique (for details, see Edwards, Hobbs \& Manchester 2006), a ``pulsar timing model" is developed that describes the pulsar, its companions and the propagation of the pulses from the pulsar to the Earth.  The model predicts the pulse times of arrival (ToAs) in an inertial reference frame.  In all current pulsar timing experiments this reference frame is taken to be the solar system barycentre. The barycentric ToAs are then compared with the predictions of the timing model and the differences are identified and termed the ``pulsar timing residuals''. As GWs are not, by default, included in the timing model, they  will induce timing residuals.

It is thought that observations of millisecond pulsars will lead to the direct detection of GWs  \citep{Sazhin1978,Detweiler1979,Jenet2005}.
A promising class of GW source potentially detectable through pulsar timing is a stochastic background, which could be generated by
an ensemble of individually unresolvable inspiraling SMBHBs scattered throughout the Universe (e.g., Sesana, Vecchio \& Colacino, 2008). Such GWs are inherently stochastic and will induce correlated noise-like structure in the timing residuals.  Other classes of GW sources are continuous waves generated by relatively nearby and massive inspiraling SMBHBs (Sesana et al. 2009, Ravi et al. 2012) and the GWM events (e.g.,  Seto 2009, Pshirkov et al. 2010, van Haasteren \& Levin 2010, Cordes \& Jenet 2012) that are the focus of this paper.  Such GW signals are deterministic and can be included in pulsar timing models.



Several observing programs have now been started with the goal of observing a large number of pulsars with sufficient precision to detect GW signals (Jenet et al. 2009; Ferdman et al. 2010; Manchester et al. 2013). Such  projects are known as pulsar timing arrays (PTAs) \citep{Romani1989,Foster1990} and currently three exist. The North American PTA (NANOGrav; McLaughin 2013) was formed in 2007 and carries out observations with the Arecibo and Green Bank telescopes.  The European Pulsar Timing Array (EPTA; Kramer \& Champion 2013) was established in 2004 and includes telescopes in England, France, Germany, the Netherlands and Italy.  For this paper we make use of data from the PPTA project (Manchester et al. 2013) which commenced in 2004 and uses the 64-m diameter Parkes radio telescope. Parkes observations have been used to place an upper bound on a stochastic background of GWs (Shannon et al., 2013) and to search for GW signals from individual, non-evolving, supermassive black hole binaries (Zhu et al. 2014).

In this paper
we focus on the GWM phenomenon. Seto (2009), van Haasteren \& Levin (2010) and Pshirkov et al. (2010)
 have independently shown
that pulsar timing arrays would be sensitive to sufficiently strong GWM  events.
GWM events passing a pulsar will lead to a glitch event in the timing residuals of only
that pulsar  (Cordes \& Jenet, 2012) and may be indistinguishable from a rotational glitch\footnote{Pulsar glitch events lead to a sudden frequency increase. Sometimes this is followed by an exponential relaxation.  It is also often found that sudden changes in the spin-down rate occur at the time of the glitch which again, may or may not, relax after the event. GWM events simply lead to a change in the pulse frequency. Depending upon the pulsar-Earth-GWM angle this may be positive or negative.}.  GWM events passing the Earth will lead to simultaneous glitch events that are potentially detectable in the timing residuals of multiple pulsars in the array. The size and sign of the glitches will depend upon the
angle between the source, Earth and pulsar in a quadrupolar fashion. Thus such events can be separated from rotational glitches in individual pulsars.

Along with GWs, many other physical phenomena are not included in the timing model and will also induce
timing residuals that may mask the signals of interest.  These include errors in the terrestrial time standard
(see e.g., Hobbs et al. 2012), errors in the solar system ephemeris (e.g.,
Champion et al. 2010), and uncorrected dispersion measure variations (see e.g.,
Keith et al. 2013).   Pulsars are also known to exhibit intrinsic variations in the timing residuals which include stochastic spin noise (e.g., Shannon \& Cordes 2010, Hobbs et al. 2010) and glitch events (e.g., Yu et al. 2013, Cognard \& Backer 2004).

In this paper we try to answer the following four questions:

\noindent (i)  How can we detect GWM signals in pulsar data sets?

\noindent (ii) Do GWM signals exist in the Parkes Pulsar Timing Array data set?

\noindent (iii) If no signal is detected then what is the maximum rate estimate of such GWM events?

\noindent (iv) What are the astrophysical implications  of this bound on the GWM amplitude?

In Section 2, we describe
the  observations used in this analysis. In Section 3, we describe the
GWM signal. In Section 4 we present our detection algorithm  for
searching and limiting the GWM signal and answer question (i).  In Section 5, we apply our algorithm
to the PPTA data and present the results, answering question (ii). In Section 6 we discuss our results and their astrophysical implications.  This leads to answers to questions (iii) and (iv). In the appendix we describe updates made to the software package \textsc{tempo2} needed for this work.

\section{Observations}
 \begin{figure*}
\includegraphics[height=21cm]{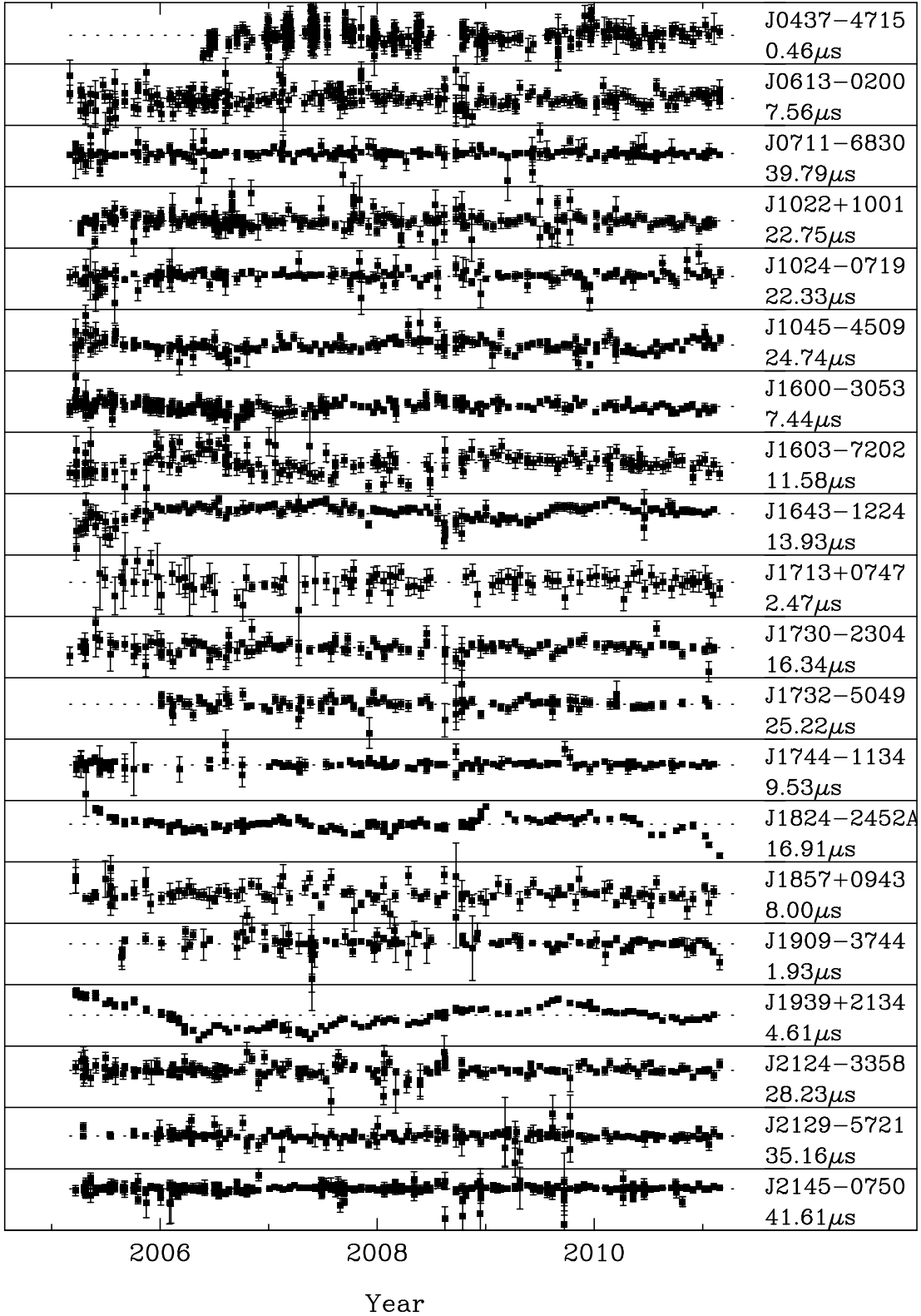}
 \caption{The post-fit timing residuals for the PPTA data set.
 The dashed, horizontal lines indicate zero residual. The pulsar name and
 the range of the timing residuals are
 labeled on each subplot. \label{fg:residuals}}
 \end{figure*}

We use the initial Parkes Pulsar Timing Array (PPTA) data set (Manchester et al. 2013).  This data set is available from the Commonwealth Scientific and Industrial Research Organisation (CSIRO) data archive\footnote{http://data.csiro.au} and has  a Digital Object Identifier (DOI) of 10.4225/08/534CC21379C12\footnote{Accessible from the permanent link \url{http://dx.doi.org/10.4225/08/534CC21379C12}}. The raw observations that made up that data release are also available from the same website as part of the Parkes pulsar data archive
(Hobbs et al. 2011).

The data set includes regular
observations of 20 millisecond pulsars at intervals of 2-3 weeks
between the years 2005 and 2011\footnote{The Manchester et al. (2013) paper also describes an ``extended data-set" that includes earlier observations.  These earlier data cannot be corrected for dispersion measure variations and so are not used in this work.}.
For each pulsar, ToAs for the band that has the lowest over-all rms timing
residuals after the data have been corrected for dispersion measure variations (Keith et al. 2013) have been selected.
All the observations were performed with the  Parkes 64-m radio telescope with typical integration times of 1\,hr.

Timing residuals were formed using the \textsc{tempo2} software package (Hobbs et al. 2006) making
use of the JPL DE421 Solar System ephemeris (Folkner et al. 2008) and referred to terrestrial time as
 realised by the Bureau International des Poids et Mesures\footnote{\url{http://www.bipm.org}} (BIPM2011).  The post-fit timing residuals for the 20 pulsars are shown in Figure~\ref{fg:residuals}.

The first three columns in Table~\ref{tb:params} provide, for each pulsar, its name, pulse period and dispersion measure.  Pulsar timing data sets vary significantly and we currently do not have a simple way to quantify the quality of different data sets.  Usually the weighted rms of the timing residuals is used.  We present this value, $\sigma_{\rm w}$, in column 4.  However, we note that pulsars scintillate and therefore some ToAs can have much larger uncertainties than other ToAs for the same pulsar.  We therefore also present the unweighted rms timing residuals, $\sigma_{\rm uw}$, in column 5.   Both of these statistics are affected by any non-white noise process in the data and so the uncertainties on individual ToAs are often significantly lower than the rms values.  To quantify this, we give, in column 6, the median ToA uncertainty for each pulsar, $m_{\rm orig}$, as measured during the ToA determination procedure.   Below, we show that additional white noise, which is not described by the ToA uncertainties, may also be present.  After correction for such noise, the median uncertainty will be increased.  This corrected median ToA uncertainty, $m$, is listed in column 7 of the table.  The remaining three columns give the number of ToAs (N$_{\rm obs}$), the dates of the first and last observations (as Modified Julian Dates, MJDs) and the data span respectively.

\begin{table*}
\caption{The basic parameters for the PPTA data set}\label{tb:params}
\begin{tabular}{lrrrrrrrrr}
\hline
\multicolumn{1}{l}{PSR J} & \multicolumn{1}{c}{Period} & \multicolumn{1}{c}{DM} & \multicolumn{1}{c}{$\sigma_{\rm w}$} & \multicolumn{1}{c}{$\sigma_{\rm uw}$} & \multicolumn{1}{c}{$m_{\rm orig}$} & \multicolumn{1}{c}{$m$} & \multicolumn{1}{c}{N$_{\rm obs}$} & \multicolumn{1}{c}{Range} & \multicolumn{1}{c}{Span} \\
       &\multicolumn{1}{c}{(ms)}   & \multicolumn{1}{c}{(cm$^{-3}$pc)} & \multicolumn{1}{c}{($\mu$s)}   & \multicolumn{1}{c}{($\mu$s)}     & \multicolumn{1}{c}{($\mu$s)} & \multicolumn{1}{c}{($\mu$s)} & & \multicolumn{1}{c}{(MJD)} & \multicolumn{1}{c}{(yr)} \\
\hline
J0437$-$4715 & 5.757 & 2.64 & 0.071 & 0.074 & 0.03 & 0.06 & 475 & 53880---55618 & 4.8 \\
J0613$-$0200 & 3.062 & 38.78 & 1.144 & 1.371 & 0.88 & 1.06 & 218 & 53431---55619 & 6.0 \\
J0711$-$6830 & 5.491 & 18.41 & 1.341 & 4.367 & 2.41 & 2.44 & 212 & 53431---55619 & 6.0 \\
J1022+1001 & 16.453 & 10.25 & 1.901 & 2.239 & 0.94 & 1.62 & 211 & 53468---55618 & 5.9 \\
J1024$-$0719 & 5.162 & 6.49 & 1.117 & 2.939 & 1.71 & 1.74 & 175 & 53431---55620 & 6.0 \\
J1045$-$4509 & 7.474 & 58.15 & 2.480 & 3.249 & 2.13 & 2.13 & 183 & 53451---55620 & 5.9 \\
J1600$-$3053 & 3.598 & 52.33 & 0.724 & 0.837 & 0.50 & 0.65 & 237 & 53431---55598 & 5.9 \\
J1603$-$7202 & 14.842 & 38.05 & 2.446 & 2.616 & 1.00 & 1.66 & 168 & 53431---55618 & 6.0 \\
J1643$-$1224 & 4.622 & 62.41 & 1.593 & 2.024 & 0.67 & 0.81 & 133 & 53453---55598 & 5.9 \\
J1713+0747 & 4.570 & 15.99 & 0.514 & 0.535 & 0.22 & 0.47 & 98 & 53533---55619 & 5.7 \\
J1730$-$2304 & 8.123 & 9.62 & 1.679 & 2.289 & 1.19 & 1.77 & 130 & 53431---55598 & 5.9 \\
J1732$-$5049 & 5.313 & 56.83 & 2.355 & 3.189 & 2.09 & 2.32 & 102 & 53725---55581 & 5.1 \\
J1744$-$1134 & 4.075 & 3.14 & 0.360 & 0.885 & 0.38 & 0.51 & 132 & 53453---55598 & 5.9 \\
J1824$-$2452A & 3.054 & 120.50 & 2.324 & 2.224 & 0.48 & 0.93 & 178 & 53519---55619 & 5.8 \\
J1857+0943 & 5.362 & 13.30 & 0.817 & 1.386 & 1.09 & 1.07 & 121 & 53431---55598 & 5.9 \\
J1909$-$3744 & 2.947 & 10.39 & 0.118 & 0.247 & 0.16 & 0.17 & 125 & 53605---55618 & 5.5 \\
J1939+2134 & 1.558 & 71.02 & 0.806 & 0.888 & 0.14 & 0.21 & 139 & 53451---55598 & 5.9 \\
J2124$-$3358 & 4.931 & 4.60 & 1.917 & 3.633 & 2.17 & 2.21 & 186 & 53431---55618 & 6.0 \\
J2129$-$5721 & 3.726 & 31.85 & 0.873 & 3.709 & 2.24 & 2.27 & 182 & 53477---55618 & 5.9 \\
J2145$-$0750 & 16.052 & 9.00 & 1.083 & 3.549 & 1.24 & 1.29 & 482 & 53431---55618 & 6.0 \\
\hline
\end{tabular}
\end{table*}

All pulsar data sets have a number of properties that make searching for the GWM effect challenging. Although the data span is similar for each pulsar in the PPTA data set, the data sampling is irregular and is not the same for each pulsar. The ToA uncertainties are time-variable and tend to decrease with time as new receivers are commissioned and/or wider bandwidths become available.  The variability of the residuals is quite different between pulsars (depending on, for instance, the scintillation properties of that pulsar) and low-frequency fluctuations in the residual time series are evident in many pulsars.  Such low-frequency variations are probably dominated by stochastic spin noise, but can also be caused by imperfect correction for dispersion measure variations or slight errors that arise when combining data sets from different observing instruments.

The ToAs are estimated by fitting a template pulse shape to the observations and the errors in the TOAs are estimated from the mean squared difference between the template and the observed pulse shape (Taylor 1992). The errors thus include radiometer noise and also any factors that change the pulse shape. The latter include changes in pulse shape with observing frequency which we do not include in our timing model.
All pulsars are affected by noise intrinsic to the emission known as stochastic wideband impulse-modulated self noise or ``jitter" (Os{\l}owski et al. 2011, 2013, Shannon \& Cordes 2012). For some pulsars, this additional noise significantly affects the pulse shape and violates the assumptions of the Taylor (1992) ToA estimation algorithm yielding biased ToAs with underestimated uncertainties.
There are also variations in the arrival times caused by errors in calibration (and many other processes).  The net effect is that only part of the white noise is described by the estimated uncertainty in the standard ToA determination. We have used the \textsc{efacEquad} plugin to the \textsc{tempo2} software package in order to rescale the ToA uncertainties so that they better represent the observed scatter in the residuals (Appendix A contains more details about this plugin). The plugin estimates and removes the red noise then scales the ToA uncertainties using \textsc{efac} and \textsc{equad} as defined below, choosing the values which best match the normalised residuals to a gaussian probability density using a Kolmogorov-Smirnov test. The scaled ToA uncertainty $\sigma_s$ is related to the original uncertainty $\sigma$ by:
\begin{equation}
\sigma_s^2 = \left(\sigma^2 + {\rm \textsc{equad}}^2\right) \times \rm{\textsc{efac}}^2.
\end{equation}
Different \textsc{efac} and \textsc{equad} values are obtained, in general, for each pulsar and each backend. Typical \textsc{efac}s are $\sim 1-2$ and \textsc{equad}s $\sim0-2 \mu s$.

\begin{figure}
 \includegraphics[height=8cm,angle=-90]{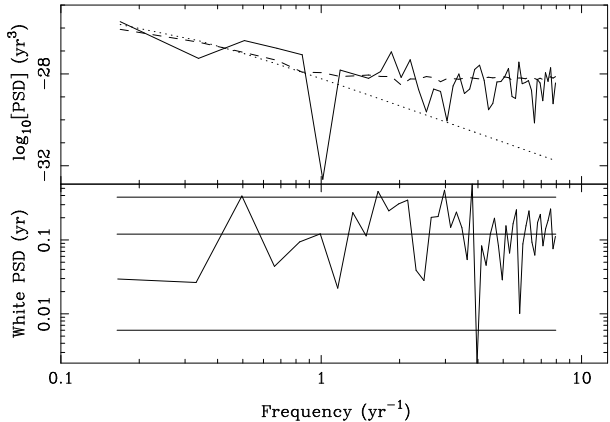}
 \caption{(top panel) Power Spectrum Density (PSD) of the timing residuals for PSR~J1643$-$1224 (solid line).  A model of the red noise is indicated as the dotted line.  The dashed line is the mean power spectrum of 100 simulations of the white and red noise. (lower panel) Power spectrum for the whitened residuals of PSR~J1643$-$1224 (solid line).  The expected mean and $\pm2\sigma$ confidence intervals are shown as horizontal lines. \label{fg:spectra_1643}}
 \end{figure}

Many of the pulsar data sets also exhibit red noise for which no prior estimate is available. An estimate of the covariance matrix of this red noise must be obtained to optimize the timing analysis and ultimately the detection process.
For each pulsar we used the \textsc{spectralModel} plugin to look for evidence of non-white noise.  When such noise was detected, we obtained a self-consistent estimate of the covariance matrix of the low-frequency noise using the iterative procedure discussed by Coles et al. (2011). An initial estimate of the red noise spectrum was obtained and a model fitted to it. This was used to estimate the covariance matrix of the red noise. The white noise component of the variance was added to the diagonal to obtain the complete covariance matrix and which used to estimate the power spectrum using a generalised least squares fit. An improved model was fitted to this power spectrum and the process iterated until a self-consistent solution was obtained.
The following simple red noise model is adequate for our data sets and has some physical justification (see, e.g., Shannon et al. 2013; Melatos \& Link 2014):
\begin{equation}
P_r(f) = {P_0}{\left[1+\left({f}/{f_c}\right)^2\right]^{-\alpha/2}}.
\end{equation}
An example for a representative pulsar (PSR~J1643$-$1224) is shown in the top panel of Figure~\ref{fg:spectra_1643}.  Here the solid line is the power spectral density estimate of the data and the dotted line is the analytical model. The dashed line is the average of 100 spectra obtained from simulated data sets. These data sets had the same white noise and observing cadence as the original data along with red noise with the spectral properties defined by the analytical model.

The objective of this noise modelling is to find a linear transformation that whitens and normalises the residuals. The \textsc{efac}s and \textsc{equad}s used to model the white noise are not unique, nor is the red noise spectral model. The test we use for a satisfactory linear transformation is that the power spectrum of the whitened residuals fits
within the $\pm2\sigma$ error bars for a  random variable with two degrees of freedom. We use the Lomb-Scargle algorithm to estimate this spectrum because it is unbiased for a truly white time series. The power spectrum for PSR~J1643$-$1224 after whitening is shown in the lower panel of Figure~\ref{fg:spectra_1643}. We have included the parameters of the red noise models we used in Table~\ref{tb:redPar} so that others can duplicate our analysis. The \textsc{efac} and \textsc{equad}s are tabulated in Appendix B.  We note that the development of these noise models is subjective and discuss the implications of this in Section~\ref{sec:noiseModels}.


\begin{table}
\small
\caption{The red noise parameters for PPTA data sets.}\label{tb:redPar}
\begin{tabular}{llll}
\hline
PSR        &  $\alpha$  & $P_0$ (yr$^3$) &$f_c$ (yr$^{-1}$)         \\
\hline
  J0437$-$4715 &  3 &4.3 $\times 10^{-29}$ &0.2   \\
  J0613$-$0200 &  5 &5.5 $\times 10^{-28}$ &0.5  \\
  J1022$+$1001 &  3 &1.8 $\times 10^{-27}$ &0.5   \\
  J1024$-$0719 &  4 &3.9 $\times 10^{-27}$ &0.4   \\
  J1045$-$4509 &2.5 &2.1 $\times 10^{-26}$ &0.2   \\
  J1600$-$3053 &  4 &3.8 $\times 10^{-27}$ &0.2    \\
  J1603$-$7202 &  3 &2.3 $\times 10^{-27}$ &0.2  \\
  J1643$-$1224 &  4 &1.5 $\times 10^{-26}$ &0.2   \\
  J1713$+$0747 &  4 &4.1 $\times 10^{-28}$ &0.2 \\
 J1824$-$2452A &  4 &1.3 $\times 10^{-25}$ & 0.2   \\
  J1939$+$2134 &2.5 &2.8 $\times 10^{-27}$ &0.2   \\
  J2129$-$5721 &4 &3.5 $\times 10^{-27}$   &0.2   \\
\hline
\end{tabular}
\end{table}


\section{The GWM Signal in Pulsar Timing}
\label{sec:signal}
In the example of two coalescing equal-mass black holes, the amplitude of the GWM signal grows rapidly.
The growth timescale
of the metric change is  $\sim$ 10$^{4} {\rm s} ({\rm M}/10^8{\rm M}_\odot)(1 + z)$ (van Haasteren \& Levin 2010),
where $z$ is the redshift of the source and ${\rm M}$ is the mass of each black hole (the black holes are assumed to have equal mass). This is short compared with typical observation intervals for existing PTAs (which is normally one observation every 2--3 weeks).  We assume such growth timescales for all GWM sources, and treat  the signal as a discrete jump of the metric propagating through space.

 We model the GWM signal as a step-function
\bea \label{eqn:gwm}
  h_{+}(t) = h^{\rm mem}\Theta(t-t_0),~~h_{\times}(t) = 0,
\ena
where $t_0$ is the time the GWM signal reaches the observer on Earth. Note that Favata (2009) showed that the definition of ``plus" and ``cross" polarisation can be such that the memory signal only causes a shift in
the amplitude of the ``plus" polarisation for systems in a circular orbit.
The function $\Theta(t)$ is the Heaviside
step function \bea \Theta(t) = \left\{\begin{array}{l} 0,~~t\leq0
\\ 1,~~t>0 \end{array} \right. \label{Theta1} \ena

Determining the exact functional form for the characteristic strain is challenging as memory is produced predominantly in the final moments of a merger when Einstein's  equations are not analytically solvable.  The most detailed calculations for the size of the burst were given by Favata (2009).  In Equation 1 of Madison, Cordes \& Chatterjee (2014) a detailed prediction for the amplitude of the memory event that depends upon the black hole masses, the inclination angle of the orbit just prior to merger and the source distance is given.
Cordes \& Jenet (2012) provide a simple way to obtain order-of-magnitude estimates of the signal strength from
\begin{eqnarray}
 h^{\rm mem}\sim
 5\times10^{-16}\left({\rm \mu}/{10^8~{\rm M}_{\odot}}\right)\left({1~\mathrm{Gpc}}/{\rm D}\right)
 \label{eq:pshirkov}
\end{eqnarray}
where ${\rm D}$ is the distance between the Earth and the SMBHB and $\mu$ is its reduced mass.

 For a single pulsar, the fractional frequency change caused by a
plane gravitational wave (Estabrook \& Wahlquist 1975; Hellings \& Downs 1983) is:
  \begin{eqnarray}
	  {\delta \nu(t) / \nu}&=&B(\theta, \phi) \left[h(t)-h(t-r/c(1 + \cos\theta))\right]
 \label{eqn:dn}\end{eqnarray}
 and
  \begin{eqnarray}
    B(\theta,\phi) &=&\frac{1}{2} \cos (2\phi) \left( 1-\cos\theta\right).
   \end{eqnarray}
Here $c$ is the vacuum speed of light, $r$ is the distance from the Earth to the pulsar,
$\theta$ is the angle
between the direction from the observer to the pulsar and the direction of GW
propagation, $\phi$ is the angle between the wave's principal
polarisation and the projection of the pulsar direction onto the plane perpendicular to
the propagation direction, and $h(t)$ is the strain of the gravitational-wave at the
observer's location.  For the analysis described in this paper
we treat the effects of all other GW signals on pulsar data as noise, and so h(t) = h$^{\rm mem}$(t).

 \begin{figure}
\includegraphics[angle=-90,width=6cm]{plot_3d.ps}
\includegraphics[width=6cm]{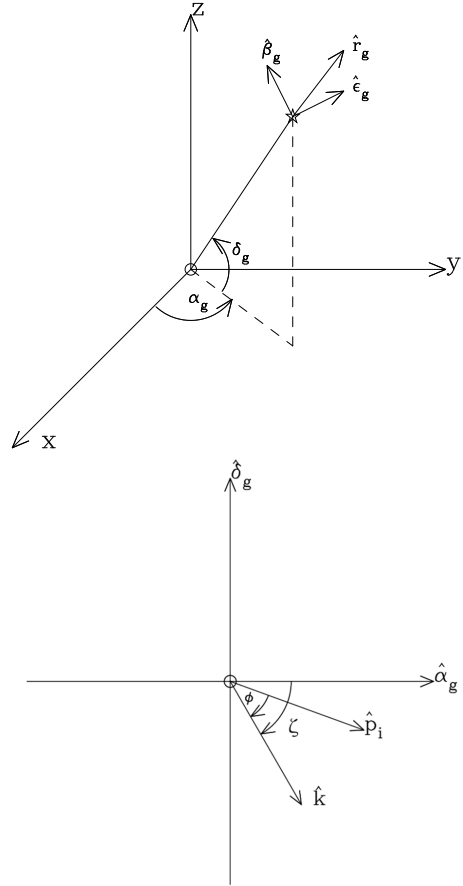}
\caption{The geometry used to describe the GWM emission.  The top panel has the Earth at the centre and the GWM source indicated by the star symbol.  The lower panel represents the GWM emission coming out of the page. 
}
\label{fg:geometry}
\end{figure}

From Equations~\ref{eqn:gwm} and \ref{eqn:dn} (note that this is the same as equation 4 in van Haasteren \& Levin 2010)
\begin{equation}
    {\delta \nu(t) / \nu}=h^{\rm mem} B(\theta, \phi)\times \left[ \Theta(t-t_0)-\Theta(t-t_1)\right].
    \label{deltanu1}
  \end{equation}
   where  $t_1 = t_0 + (r/c) (1 + \cos\theta)$ is the time that the GWM event passed the pulsar.
  Therefore, the memory event  causes two
  pulse frequency jumps with the same amplitude, but with the opposite sign, separated by
  the time interval $t_0 - t_1$.

The pre-fit\footnote{Note that all residuals analysed have already been fitted to an initial model.  Here we use the term pre-fit to denote the residuals before fitting for the GWM event.}  timing residuals for a GWM event passing the Earth at $t=t_0$ are the integral of $\delta\nu/\nu$:
  \begin{equation}
r(t)_{\rm prefit} = h^{\rm mem} B(\theta,\phi)(t-t_0)\Theta(t-t_0).
  \end{equation}
The induced timing residuals for a particular pulsar therefore depend upon $h^{\rm mem}$, $t_0$, the sky position of the GWM source defined in equatorial coordinates ($\alpha_g$,$\delta_g$), the coordinates of the pulsar ($\alpha_p$,$\delta_p$) and the principal polarisation angle ($\zeta$) for the GWs.  These angles are shown graphically in Figure~\ref{fg:geometry}.  The top panel is centred on the Earth, with the north celestial pole in the z-direction. The GWM source is indicated using a star symbol.  $\hat{r_g}$ is a unit vector pointing in the direction of the source.
$\hat{\beta}_g$ is a unit vector perpendicular to the source direction in the source-Earth-z plane and $\hat{\epsilon}_g \times \hat{\beta}_g $ = $\hat{r}_g$.
 For the bottom panel we assume that the source is centred in the diagram and the GWM propagation is out of the page.  $\hat{p}_i$ represents a vector pointing to the i'th pulsar projected on the plane perpendicular to the GWM propagation. $\hat{k}$ is the principal polarisation vector for the GWM and $\phi$ has been defined after Equation 7.

 \begin{figure}
\centerline{\includegraphics[width=5.5cm,angle=-90]{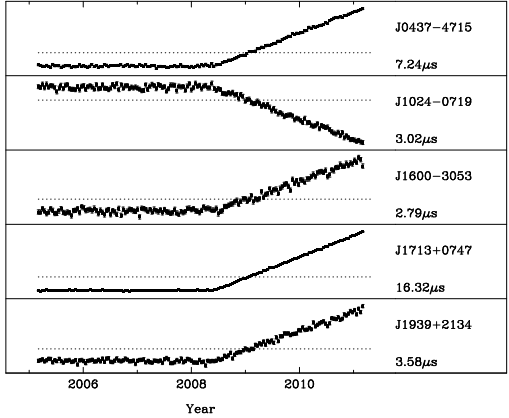}}
 \caption{ Simulated timing residuals for five pulsars.  The pulsars are affected by white noise and a GWM event that occurred at the centre of the data span. No pulsar parameters have been fitted to the timing residuals. The value underneath the pulsar's name gives the range of the timing residuals for each pulsar.
  \label{fg:simResiduals1}}
 \end{figure}

  \begin{figure}
 \centerline{\includegraphics[width=5.5cm,angle=-90]{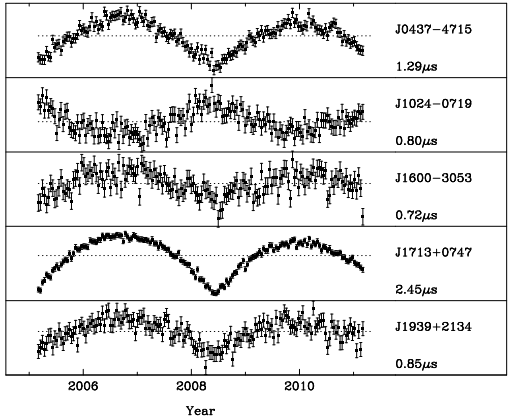}}
 \caption{As in Figure~\ref{fg:simResiduals1}, but each pulsar's pulse frequency and its derivative have been fitted and post-fit timing residuals are shown. The value underneath the pulsar's name gives the range of the timing residuals for each pulsar.
 \label{fg:simResiduals2}}
 \end{figure}

We have incorporated the effect of a GWM event into the \textsc{tempo2} timing model. This allows us to include the GWM in a fit and to simulate residuals (or ToAs) that include such an event.  The new timing model parameters are $(h^{\rm mem},t_0,\alpha_g,\delta_g,\zeta)$. 
\textsc{Tempo2} uses a linear least-squares-fitting algorithm. If the GWM epoch, polarisation angle and source position were known, it would be possible to fit for the amplitude of the GWM source as part of the standard \textsc{tempo2} timing fit.  However, if these parameters are not known then a non-linear fitting routine is needed to determine their values.

We have found this parameterisation convenient for simulating timing residuals affected by a GWM signal.  However, in order to search for the GWM events we have found it useful to provide a second parameterisation of the GWM effect within \textsc{tempo2}.  In this parameterisation we describe the GWM using two orthogonal components, $A_1$ and $A_2$ where $A_1 = h^{\rm mem} \cos (2 \zeta)$ and $A_2 = h^{\rm mem} \sin (2 \zeta)$. This formulation has the advantage that $A_1$ and $A_2$ enter the timing model linearly and can be fitted with linear least squares.
We emphasise that, even with this parameterisation, the position of the source and epoch of the event cannot be obtained using a linear fitting routine. We therefore fit for $A_1$ and $A_2$ at a grid of points for every possible sky direction and epoch.

As an example, we show in Figure~\ref{fg:simResiduals1} simulated timing residuals for five pulsars.  These timing residuals were formed by determining ToAs that are exactly predicted (i.e., yield zero residual; see Hobbs et al. 2009) by a timing model that included a GWM event (with $h^{\rm mem} = 3 \times 10^{-13}$, $\alpha_g = 21^{\rm h}51^{\rm m}, \delta_g = -30.3^\circ$) for each pulsar. We subsequently added 100\,ns of Gaussian white noise to each observation (an observing cadence of 14 days was assumed).  The resulting residuals, shown in the figure, were obtained from the initial pulsar timing models that did not include the GWM event.  The GWM event is clearly seen in the centre of the data span.  As expected, the size of the induced timing residuals depend upon the pulsar position.  The same data set is shown in Figure~\ref{fg:simResiduals2}  after fitting for the pulsars' pulse frequency and frequency derivative parameters.

\section{The Detection Algorithm}
\label{sec:algorithm}

The response of each pulsar to a single GWM burst is completely determined by specifying the source parameters: $A_1$, $A_2$, position, and epoch.  For a given position and epoch, we jointly fit $A_1$ and $A_2$ with the pulsar parameters (spindown, astrometry, orbital configuration, etc.) by adding the GWM response to the \textsc{tempo2} timing model and minimizing the whitened timing residuals.  We use the algorithm described in Coles et al. (2011) to account for the correlations in the pre-fit timing residuals caused by unmodelled red noise. To determine the sky position and epoch we search over a regular 3-dimensional grid whose spacing we describe below.

At each position and epoch the \textsc{tempo2} fit returns the parameter vector $\vec{A} = [A_1; A_2]$ and their covariance matrix, $\mathbf{C_o}$.  From these, we require a detection statistic which provides an optimal estimate of the amplitude of the GWM.  We can then use that statistic to locate the GWM in the grid of possible positions and epoch.  While this approach is not as computationally efficient as a non-linear fit, it provides an opportunity to study the statistics of the noise by examining the response over the entire three-dimensional grid.

\subsection{A Detection Statistic}

$A_1$ and $A_2$ can be viewed as the GWM amplitude modified by the response of our ``detector''.  If the pulsars in the array were distributed uniformly in position and brightness, $A_1$ and $A_2$ would be orthogonal (independent) and equally sensitive.  For this ideal case, $D\equiv A_1^2 + A_2^2$ is an optimal detection statistic.

The real non-ideal array, however, yields correlated $A_1$ and $A_2$ with differing sensitivities.  To recover the optimal detection statistic we must determine $\mathbf{U}^{-1}$, the matrix of transformation that whitens and normalises $\vec{A}$, i.e., $\vec{A}_w = \mathbf{U}^{-1} \vec{A}$.  The components of the result, $\vec{A}_w$, will be two uncorrelated random variables with unit variance. This reduces the problem to one for which we know the optimal solution is $D = A_{w1}^2 + A_{w2}^2$. This is analogous to the way we use the Cholesky decomposition to deal with red noise in \textsc{tempo2} (see Coles et al. 2011). The solution is given by
\begin{eqnarray}
D &=& \vec{A}_w^t \vec{A}_w    ~~ = ~~ \vec{A}^t \mathbf{C_o}^{-1} \vec{A},
\end{eqnarray}
where the superscript $t$ indicates transposition.

 If $\mathbf{C_o}$ is exact then, in the absence of a GWM signal, $D$ is the sum of the squares of two unit variance gaussians and thus follows a $\chi^2$ distribution with two degrees of freedom. We note that the mean and standard deviation of such a distribution both equal two, its probability density is exponential $p(D) = (1/2)\exp{(-D/2)}$, and its cumulative probability is
$c(D) = 1 - \exp{(-D/2)}$. If one chooses a detection threshold, $D^*$, then the false alarm probability is
$1 - c(D^*) = \exp{(-D^*/2)}$. For a 5\% false alarm probability, $D^* = 6$.

Two factors prevent the actual statistic from following the ideal distribution.  First, we estimate $\mathbf{C_o}$ from the data, and the statistical uncertainty in the correction of $\vec{A}$ to $\vec{A}_w$ increases the variance and biases $D$.  As the observation span increases and $\mathbf{C_o}$ becomes better characterised, the distribution of $D$ approaches $\chi^2$.  Second, as discussed earlier, we have not actually fitted optimised noise models to each pulsar. Each pulsar noise model requires five parameters and we are reluctant to fit an additional 100 parameters to the data set and reduce the degrees of freedom by a corresponding amount. An error in the noise model will not bias the parameters $A_1$ or $A_2$, but it will alter the estimate of $\mathbf{C_o}$, biasing $D$.  In both cases, as long as the shape of the distribution is unchanged, the expected false alarm probability can be recovered by renormalising $D$ such that its mean is two.

Our statistical analysis must also account for the requirement to search over a grid of possible positions and epochs.  We expect detectable GWM events to be extremely rare (see Section~\ref{sec:predict}), and so we adopt $D_{\rm max}$, the maximum $D$ in a search over epoch and position, as our final detection statistic.  Because the angular and time resolution of our array is modest, only a handful of epochs and positions are independent, and we must carry out simulations to determine the false alarm probability of $D_{\rm max}$.

\subsection{Search over possible sky positions and GWM arrival times}

\begin{figure}
\centerline{\includegraphics[width=5.5cm,angle=-90]{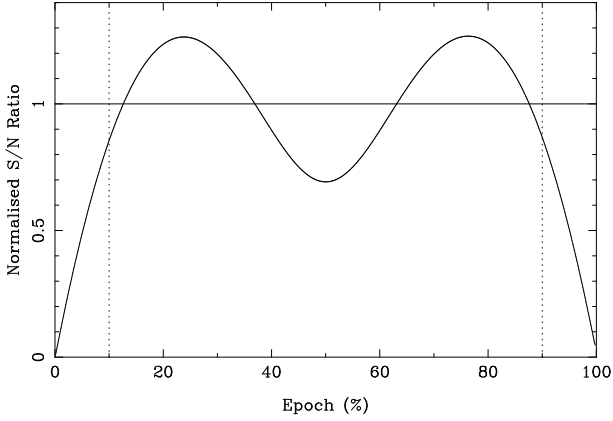}}
\caption{
\label{ramp}
The normalised S/N ratio for detection of a GWM signal in white noise as a function of the epoch of the event (calculated by fitting the amplitude of a ramp function that occurs at the specified epoch and then subsequently dividing the value by its uncertainty). The S/N ratio is normalised to the average over the range 10\% to 90\% of the observing interval.}
\end{figure}

The array sensitivity is nonuniform in both position and epoch.  In particular, sensitivity to GWM events drops to zero at the edges of the observational span, as shown in Figure~\ref{ramp}, which depicts the sensitivity for equally spaced observations with regular white noise.  This is discussed by van Haasteren \& Levin (2010) and our results concur with theirs.  Accordingly, we restrict our search over epoch to the central 80\% of the observing span, shown in the figure with a dotted line representing the region with roughly constant sensitivity.

The actual detection statistic, then, is $D_{\rm max}$ the maximum $D$ over the central 80\% of the observing interval and over all sky positions. We compute $D$ over $N_e$ points in epoch and $N_s$ points on the sky where $N_e$ and $N_s$ are selected so that $D$ is heavily oversampled, i.e., there are many fewer than $N_e N_s$ independent samples of $D$. The cumulative probability for $D_{\rm max}$ is given by
\begin{equation}
\label{eq:cdmax}
c(D_{\rm max}) = \left[1 - e^{-D_{\rm max}/2}\right]^{N_{\rm tot}}
\end{equation}
where $N_{\rm tot}$ is the total number of independent points searched. If the detection criterion is $D_{\rm max} > D_{\rm max}^*$ then the false alarm probability is $1 - c(D_{\rm max}^*)$.

\begin{figure}
\centerline{
\includegraphics[width=6cm,angle=-90]{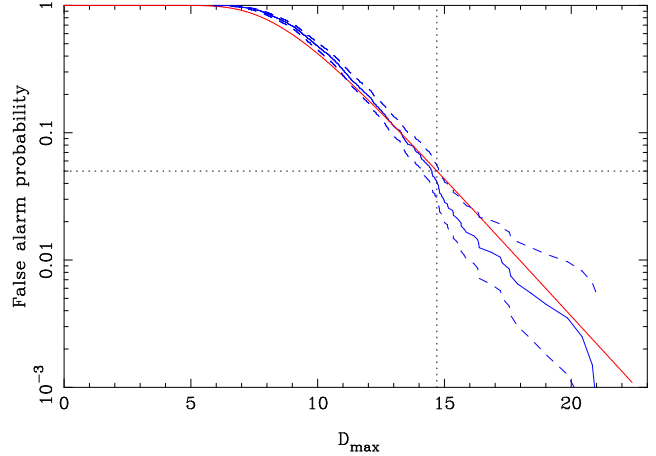}
}
\caption{
\label{false_alarm_dmax}
False alarm probability (solid blue line) $1 - c(D_{\rm max})$ obtained from 1000 simulations of the PPTA data set with both red and white noise. The blue dashed lines are the $\approx 2\sigma$ statistical errors on the measurement. The red line is the theoretical expression for $N_{\rm tot} = 80$. For a false alarm probability of 5\% (horizontal dotted line), $D_{\rm max} = 14.7$ (vertical dotted line).}
\end{figure}

To determine $N_{\rm tot}$ we performed 1000 simulations of the full PPTA data set using the red and white noise models discussed earlier. Each of the 1000 realizations was normalised, as discussed above, so the sample mean $D =2$ for each realization. We then find $D_{\rm max}$ for each realization and compute the cumulative distribution of $D_{\rm max}$, shown in Figure~\ref{false_alarm_dmax}. We determine $N_{\rm tot}$ by matching this distribution to Equation \ref{eq:cdmax}, yielding $N_{\rm tot} = 80$ and $D_{\rm max}^* = 14.7$ for a 5\% false alarm probability. In these simulations, we used $N_e = 16$ and $N_s = 34$, so the simulations were oversampled by approximately a factor of 7.

To verify that our rescaling of $D$ maintains the correct false alarm probability, we examined the 544000 simulated raw $D$ values.  While the sample mean was $\bar{D} = 2.35 \pm 0.01$, the normalised $D$ values matched the expected $\chi_2^2$ distribution within the error bars.

\section{Application to the PPTA Observations}

We applied our algorithm to the PPTA data set by calculating $D$ over a fine grid with 1034 sky positions and 150 epochs.  Our grid is therefore significantly oversampled compared with the 80 independent degrees of freedom determined above. The additional computational burden in this oversampling is insignificant for a single ``realisation" and allows us to study the correlations in detail.  We demonstrate in Section~\ref{sec:gridSampling} that statistically identical results are obtained using a smaller grid.

For each grid position we perform a global fit for $A_1$ and $A_2$ whilst simultaneously fitting for the parameters specific to each pulsar.  The mean, reduced $\chi^2$ of the timing model fits is 0.9 which indicates that the noise models were reasonably accurate, but slightly conservative, i.e., about 10\% higher in variance than the observations. The mean value of the unnormalised statistic, $\bar{D}  = 2.1$, is lower than the mean value of the simulations (2.35), but within 1$\sigma$ of it.  We therefore rescaled our values of $D$ by a small factor of $2.0/2.1 = 0.95$ for the remainder of the analysis.

\begin{figure*}
\centerline{\includegraphics[width=14cm]{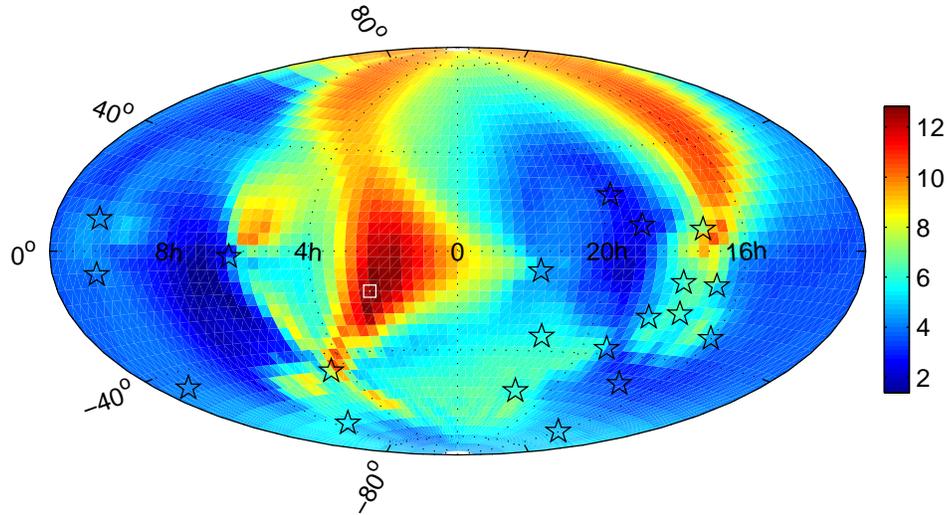}}
\caption{
\label{all_sky_2}
Normalised detection statistic, $D_e$, for the PPTA observations measured at each of 1034 sky positions. The expected maximum value of $D_e$ (see text) is $D^* = 14.7$.  The locations of the pulsars are denoted by star symbols. The white box indicates the position corresponding to the maximum value of $D_e$.}
\end{figure*}

For each sky position we calculated $D_e$, the maximum statistic determined over all epochs.  These values are shown as a function of sky position in Figure \ref{all_sky_2}. The pulsar positions are indicated on this figure using star symbols. The peak $D_e$ over the sky is $D_{\rm max} = 12.4$ (corresponding to a false alarm probability of 15\%) and occurred at MJD 54986 (corresponding to 04 June 2009) at ($\alpha_g = 02^{\rm h}24^{\rm m}, \delta_g = -15.8^\circ$).

Our value of $D_{\rm max} = 12.4$ is lower than the 5\% false alarm probability threshold of $14.7$. We therefore do not claim a detection of any GWM event in the current PPTA data set.

\subsection{Sensitivity over the sky}

\begin{figure*}
\centerline{\includegraphics[width=14cm]{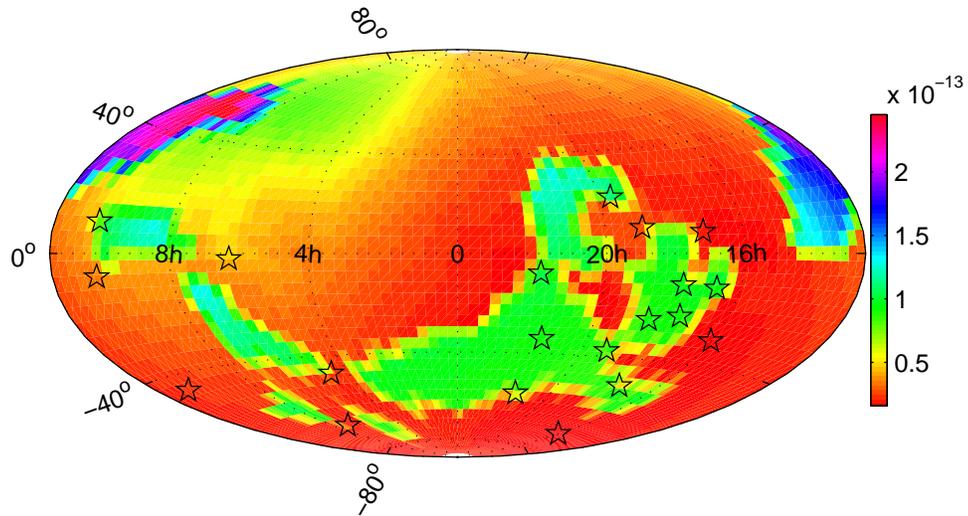}}
\caption{
\label{gwm_sens}
Sensitivity of the PPTA data set to GWM events. Note that the scale is in $10^{-13}$ leading to a maximum sensitivity of $\sim 10^{-14}$ and a worst-case sensitivity of $\sim 3 \times 10^{-13}$.}
\end{figure*}

Using our value of $D_{\rm max} = 12.4$ we can determine the sensitivity of the data set to GWMs as a function of sky position, i.e., we determine the amplitude, $h^{\rm mem}(\alpha_g,\delta_g)$, of a GWM event that would give $D = D_{\rm max}$. We assume, as discussed earlier, that the sensitivity is independent of epoch provided the epoch is restricted to the central 80\% of the observing interval. The sensitivity $h^{\rm mem}(\alpha_g, \delta_g)$ is expected to vary significantly over the sky because the pulsars are not distributed uniformly, do not have the same ToA precision nor do they have the same red noise properties.


For a GWM burst with amplitude, $h^{\rm mem}$, but unknown polarisation and epoch, the expected values
are $\langle A_1^2 \rangle = \langle A_2^2 \rangle = \left(h^{\rm mem}\right)^2/2$ and $\langle A_1 A_2 \rangle = 0$. Thus $\langle D \rangle = \left(h^{\rm mem}\right)^2 (S_{11} + S_{22})/2$ where $\mathbf{S} = \mathbf{C_o}^{-1}$. The GWM amplitude corresponding to a detection threshold $D_{\rm max}^*$ is
\begin{equation}
h^{\rm mem} = (2 D_{\rm max}^* / [S_{11} + S_{22}] )^{0.5}.
\end{equation}
$\mathbf{C_o}$ can be obtained directly from the timing model matrix and the covariance matrix of the residuals. A map of $h^{\rm mem}$ at the epoch at which the maximum $D$ occurred for each position is shown in Figure~\ref{gwm_sens}. Here we took the detection threshold $D_{\rm max}^*$ as the $D_{\rm max} = 12.4$ value obtained from the PPTA observations. One should note that the sensitivity is ``low'' where $h^{\rm mem}$ is high, i.e., it is hard to detect a GWM event with the PPTA near sky coordinates with a right ascension of 10\,hr and declination of  $+$45$^\circ$.

\subsection{Bounding}


Figure~\ref{gwm_sens} shows that we could detect events at a level of $h^{\rm mem} = 2.4\times 10^{-13}$ anywhere on the sky (with 95\% confidence) if they occurred during our five-year effective observing interval.  Our data set is sensitive to events with amplitude $h^{\rm mem} \sim 2 \times 10^{-14}$, but only over a smaller area of the sky.

As no GWM event has been detected in our data, we can determine the area of the sky that would allow a detection of a GWM event at a given amplitude and convert these values to bounds on the rate of GWM events. Assuming that the occurrence of GWM events is isotropic and follows a Poisson process then one obtains a 95\% bound on its rate parameter
\begin{equation}
\lambda(h^{\rm mem}) < \frac{-\log_e(1-p)}{T_{\rm eff}} \frac{A_{\rm sky}}{A}
\end{equation}
where $p$ is the detection probability (0.95), $T_{\rm eff}$ is the effective data span (i.e., 5\,yr), $A_{\rm sky}$ is the sky area ($4 \pi$\,sr) and $A$ is the area of the sky for which GWM events of amplitude $h^{\rm mem}$ could be detected.

 For instance, from Figure~\ref{gwm_sens} we can say that no event of $h^{\rm mem} \ge 2.4 \times 10^{-13}$ occurred during our data set.  This gives an event rate of $\lambda < 0.75~{\rm yr}^{-1}$ for events of this size. Events at a level $h^{\rm mem} < 2\times10^{-14}$ could not be detected at all, and no bound at that level can be set. The bounds on $\lambda$ obtained from the PPTA data are shown in Figure \ref{rate_bnd}.

\section{Discussion}

\subsection{Choice of grid sampling}\label{sec:gridSampling}

In the previous section we calculated $D$ over a fine grid with 1034 sky positions and 150 epochs.  This number of grid points is significantly greater than the 80 independent degrees of freedom, but does allow an easy way to make sky-maps such as those in Figures~\ref{all_sky_2} and \ref{gwm_sens}. However, in order to confirm that our astrophysical results are not significantly affected by such oversampling we re-analysed the data in an identical manner, but first using 34 sky positions and 16 epochs and secondly using 16 sky positions and 34 epochs.

Using the procedure described above, we obtained, for the first case,  $\bar{D} = 1.96$. Rescaling by $1.96/2.0$ gave $D_{\rm max} = 9.9$ (compared with $D_{\rm max} = 12.4$ from the initial grid).   This new value is also lower than the 5\% false alarm probability threshold and we therefore note that this result is consistent with that obtained from the larger grid.  To conclude this analysis we obtain a single sky-averaged bound on the event rate of 0.75${\rm yr}^{-1}$  for $h^{\rm mem} \ge 1.7 \times 10^{-13}$.  For the second case we obtain $\bar{D} = 1.95$ and, after rescaling, $D_{\rm max} = 8.4$. The sky-averaged bound on the event rate of 0.75${\rm yr}^{-1}$ for $h^{\rm mem} \ge 1.6 \times 10^{-13}$.   We therefore conclude that, as expected, our results do not significantly change with the number of grid points used.

\subsection{Sensitivity of $D_{\rm max}$ to noise models}\label{sec:noiseModels}

As we were carrying out the research for this paper, we discovered minor flaws in our initial models for the red and white noise for some pulsars. The models were corrected as the flaws became apparent, but the results we obtained with those initial models allow us to determine the effect of slightly incorrect noise models.

The penultimate noise models were the same as presented in Table~\ref{tb:redPar} except for two pulsars. For PSR~J1909$-$3744 we initially included a red noise model (with $\alpha = 4$, $P_0 = 1.2 \times 10^{-29}$ and $f_c = 0.5$) and slightly different EFAC and EQUAD values.  For PSR~J2129$-$5721 we originally used only a white noise model. Following our detection procedure using these original noise models leads to an unnormalised $D_{\rm max} = 18$ (compared with the unnormalised  $D_{\rm max} = 12.9$ using the final models). After normalising the statistic we obtained $D_{\rm max} = 14$ (compared with 12.4 with our final models).

This highlights that even with poor noise models the statistics only change slightly.  However, the poor noise models can lead to detection statistics that do slightly exceed the expected false-alarm probability and therefore any ``detection" should be treated with caution. The following checks can be carried out on such detections:
\begin{itemize}
\item Ensure that the detection does not depend solely on a single pulsar. That pulsar could exhibit a glitch event that would produce a strong response in our detection statistic.
\item The detection must exhibit the expected signature in the position and epoch parameter space. A burst will have a deterministic pattern in $A_1$ and $A_2$ which depends upon the actual ($\alpha_g,\delta_g,t_0$) of the burst.  One could then subtract this pattern from the grid of observations.  The residuals should be entirely caused by noise.  If the maximum detection statistic is not reduced then there is either another burst present (which would be unlikely, but could be tested by iteration) or the original burst is not real.
\item As long as the GWM event can be distinguished from the low frequency timing noise, the significance of any such detection would increase with longer data spans and by including extra pulsar data sets.
\item It is likely that any such GWM event would be caused by the coalescence of supermassive black holes.  It is possible that such an event may be associated with an observable signal using other telescopes (see e.g., Burke-Spolaor, 2013). As the ability to localise the source improves with time it therefore may be possible to identify the host galaxy of such an event.
\end{itemize}

\subsection{Astrophysical implications}\label{sec:predict}

Our results imply that fewer than ~0.75 events yr$^{-1}$ have occurred with h$^{\rm mem} > 10^{-13}$ and fewer than
$\sim 3$ events yr$^{-1}$ for $h^{\rm mem} > 4 \times 10^{-14}$.  In this section we compare these results with
predictions for the event rate from astrophysical models.  We note that it is extremely difficult to predict
the event rates from existing observational data.  Instead we require the use of models for SMBHB mergers and then predict the event rates and amplitudes. Those predictions also rely on assumptions of the sources, such as their distances, black hole mass ratios, etc.  Two papers have already made predictions.  Cordes \& Jenet (2012) predict an event rate of 0.4\,yr$^{-1}$ for $h^{\rm mem} > 10^{-16}$ and 0.02\,yr$^{-1}$ for $h^{\rm mem} > 10^{-15}$.  Ravi et al. (2014) used the best available observational data on black hole masses and galaxy merger rates to conclude that only $\sim 10^{-5}$ bursts yr$^{-1}$ with
$h^{\rm mem} > 5 \times 10^{-15}$ and $\sim 10^{-3}$ bursts yr$^{-1}$  with $h^{\rm mem} > 2 \times 10^{-15}$ are expected.  These predictions are both so much lower than our current bounds that 1) we are unlikely to detect such events in the near future and 2) we can make predictions on the ``time-to-detection" using a simple toy-model of the GWM event rate.  We therefore note that the same black hole binary systems that lead to a GW background (GWB) will also coalesce to form the GWM sources.  We can use this to derive a simple relation between the strength of the GWB and the GWM event rate.



\begin{figure}
 \centerline{\includegraphics[width=6cm,angle=-90]{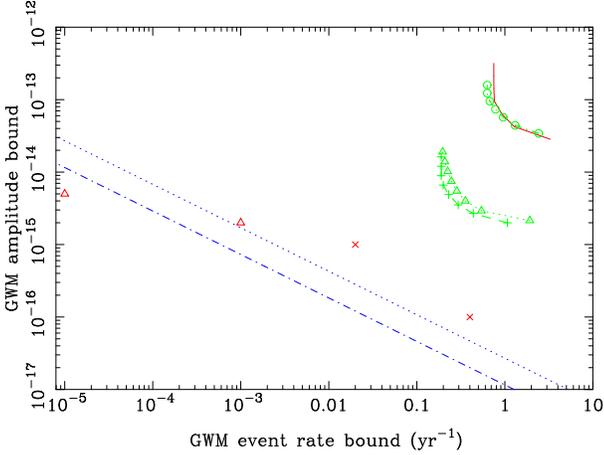}}
\caption{
\label{rate_bnd}
The solid line shows upper bounds on the event rate $\lambda$ as a function of the GWM amplitude bound using the PPTA data set. The dotted and dot-dashed lines at the bottom of the plot are predictions (detailed in the text) based on models of coalescing supermassive binary black holes. The red triangles are event rate predictions from Ravi et al. (2014). The red crosses are the predictions from Cordes \& Jenet (2012). Results of simulations are also shown as dotted or dashed green lines with symbols, with the lower set based on simulated 20-year data sets with either optimistic or pessimistic extrapolation of the red noise model (see Section 6.5 for details).}
\end{figure}

We consider a GWB that is entirely generated by a SMBHB population at a single redshift $z$ with equal-mass SMBHs of mass $M$ (corresponding to a reduced mass $\mu=M/2$). Then, following, e.g., Phinney (2001) and Sesana et al. (2008), the characteristic amplitude of the GWB at a frequency of $f_{\rm yr}=(1\,{\rm yr})^{-1}$, $A_{\rm yr}$, is given by
\begin{equation}
A_{\rm yr} = \left[f_{\rm yr}\frac{dn}{dt}\frac{dt}{df}h_{s}^{2}(\mu,\,z)\right]^{1/2},
\label{a_yr}
\end{equation}
where $dn/dt$ is the all-sky coalescence rate of SMBHBs observed at the Earth and
\begin{equation}
\frac{dt}{df}=\left[\frac{96}{5}c^{-5}\pi^{8/3}f_{\rm yr}^{11/3}(2^{4/5}GM)^{5/3}\right]^{-1}
\label{eq:dtdf}
\end{equation}
where $dt/df$ is the rate of evolution of the emitted GW frequency of a binary SMBH measured at the Earth
with $c$ as the vacuum speed of light, $G$ the gravitational constant, and
\begin{equation}
h_{s}=\sqrt{\frac{32}{5}}\frac{(2^{4/5}GM)^{5/3}}{c^{4}D(z)}[\pi f_{\rm yr}(1+z)]^{2/3}
\end{equation}
is the rms GW strain amplitude produced by a SMBHB. Re-arranging Equation~\ref{a_yr}, we obtain
\begin{eqnarray}
\label{rate}
\frac{dn}{dt} = \lambda &=& \left(6\times10^{-3}\,{\rm yr}^{-1}\right)\left(\frac{A_{\rm yr}}{10^{-15}}\right)^{2} \times \\ \nonumber
&& \left(\frac{\mu}{10^{8}M_{\odot}}\right)^{-5/3}\left(\frac{D(z)}{1\,{\rm Gpc}}\right)^{2}(1+z)^{1/3}.
\end{eqnarray}
Using Equation~\ref{eq:pshirkov} we can express Equation~\ref{rate}  in terms of $h^{\rm mem}$ for fixed $D$ or $\mu$. In Figure~\ref{rate_bnd}, we plot the event rate at different $h^{\rm mem}$ for SMBHBs at redshifts of $0.05$ (dot-dashed line) and 2 (dotted line), and for SMBHBs with reduced masses of $10^{8}M_{\odot}$. We assume $A_{\rm yr}=10^{-15}$.

For any PTA with a data span of six years, the rate bounds will lie between 0.8 and 3.5\,yr$^{-1}$ and the most constraining bound will be around 1\,yr$^{-1}$. At that rate the PPTA sensitivity  must be increased by a factor of 1700 to begin to constrain the existing theory. The current lack of detections is therefore entirely expected. Note that the observational bound will significantly improve with longer and more precise data sets and will move diagonally down and to the left in Figure~\ref{rate_bnd}. In Section~\ref{sec:expect} we describe the data sets that would be needed in order to reach the required sensitivity to make a detection of a GWM event.

We note also that although SMBH binaries are the most commonly
predicted GWM emitters detectable by pulsar timing, our bounds on GWM
rates do not only apply to memory events associated with SMBH
binaries. In this regard, our results put the strongest limits to date
on this ``gravitational wave discovery space'' as described in
\citet{cutler+14}.

\subsection{Pulsars that contribute to the bound}

The current PPTA data set contains observations of 20 pulsars.
It is interesting to consider which pulsars actually contribute to the resulting bounds on GWM events. For instance, our available observing time can be used more efficiently if we know which pulsars do not significantly affect our scientific results. To determine this we eliminated each of the PPTA pulsars from the existing array in turn, and examined the resulting rate bounds obtained for a GWM amplitude of 5$\times 10^{-14}$. In this way we ranked each pulsar by the increase in the bound ($\Delta \lambda$) that occurred when it was eliminated from the array. The nine pulsars shown in Table~\ref{tb:fewerPsr} are sufficient to obtain a bound 10\% larger than the one we actually obtained.

\begin{table}
\caption{The event rate bound for GWM amplitude 5$\times 10^{-14}$ if we eliminate a pulsar from the PPTA (compared with $\lambda = 0.60$ when all the pulsars are included)}\label{tb:fewerPsr}
\begin{tabular}{llll}\hline
Rank & PSR & $\Delta \lambda$ (yr$^{-1}$) & $\lambda$ (yr$^{-1}$) \\ \hline
1 & J1909$-$3744 & 0.34 & 0.94 \\
2 & J0437$-$4715 & 0.30 & 0.90 \\
3 & J1744$-$1134 & 0.18 & 0.78 \\
4 & J2129$-$5721 & 0.08 & 0.68  \\
5 & J1730$-$2304 & 0.08 & 0.66 \\
6 & J2145$-$0750 & 0.08 & 0.66 \\
7 & J1713$+$0747 & 0.07 & 0.65 \\
8 & J1857$+$0943 & 0.04 & 0.64 \\
9 & J1939$+$2134 & 0.04 & 0.64 \\
\hline
\end{tabular}
\end{table}

For this paper we therefore could have achieved almost the same result by only processing these pulsars.  However, we do not recommend that the PPTA significantly reduces the number of pulsars observed even though more observing time could then be spent on the ``best" pulsars.   In order to confirm a detection of either a GWM event or a GW background it will necessary to identify the correlated signals using observations of numerous pulsars.

\subsection{Expectations for the future}\label{sec:expect}

Our datasets will become more sensitive to GWM events with longer and/or improved data sets.
The Parkes data is being combined with observations from northern hemisphere telescopes as part of the International Pulsar Timing Array (IPTA) project (see e.g., Hobbs et al. 2010 and Manchester 2013). The initial IPTA data set is currently not finalised, but should lead to the best available data sets for carrying out this research. Currently 50 pulsars are being timed as part of the IPTA (see Manchester 2013).

In the more distant future, telescopes such as the Five-hundred-meter Aperture Spherical Telescope (FAST), the South African MeerKAT radio telescope and the Square Kilometre Array should produce even more sensitive data sets. Exactly how sensitive these data sets will be depends upon to what degree the pulsars will be affected by jitter noise (e.g., Os{\l}owski et al. 2013, Shannon \& Cordes 2012) and/or intrinsic timing noise (e.g., Hobbs et al. 2010, Shannon \& Cordes 2010).  Jitter noise will limit the achievable ToA precision whereas timing noise will limit the long term stability of pulsar data sets.

Our event rate bounds are currently orders-of-magnitude away from the expected event rates. We therefore consider what changes in observations would be necessary to constrain astrophysical models of the GWM rate. Longer data sets and more sensitive arrays are the obvious choices. Longer data sets are more sensitive to the signature of a GWM event in the residuals and the chance of an event occurring within the data set will increase as the data span increases. We therefore estimate the effect of extending the time span of the existing PPTA data set using simulations, and we estimate the effect of more sensitive arrays using an analytical model.

Firstly, we expand the existing PPTA data sets to have a 20-year data span assuming no significant change in the number of pulsars observed, their observing cadence or the timing precision achieved. We also assume that the existing red noise models are adequate for extrapolation to the longer time spans. We try an ``optimistic" (in terms of GW detection) extrapolation which assumes that the red noise plateaus at the current data span (i.e., we leave the corner frequency as is) and a ``pessimistic'' extrapolation in which we assume that the red noise power law continues without a corner frequency for the extended data span.	

We form simulated arrival times for each pulsar. We then calculate the sensitivity of the data set to GWM events that occurred at a time corresponding to 15\% of the data span.  As before this provides us with bounds on the event rate for a given GWM amplitude.  We overplot in Figure~\ref{rate_bnd}, a green, dashed curve that represents the bound obtained from the pessimistic extrapolation of the red noise (overlaid with triangle symbols) and the green, dotted curve that represents the bound obtained from the optimistic extrapolation (overlaid with cross symbols).  For comparison, we also plot the bound obtained using the same data span and red noise models as the actual data (green, dot-dashed line; circle symbols).  This should agree with the solid curve.  The slight difference occurs as 1) these new curves represent GWM events occurring at 15\% of the data span, as opposed to at any time within central 80\% of the data span and 2) the real data contains effects that are not perfectly modelled by the red and white noise models.

We note that adding 14 extra years does, as expected, make a significant improvement to the bounds (In Figure~\ref{rate_bnd} the bounds move downwards and to the left).  However, the use of the pessimistic red noise extrapolation only slightly reduces the bounds.  This is because many of the pulsar data sets that are highly ranked in Table~\ref{tb:fewerPsr} (such as PSRs~J1909$-$3744 and J1744$-$1134) have no red noise model (they are assumed to have white residuals).  It is likely that these pulsars do exhibit timing noise which is currently undetectable and making more realistic predictions of how the bounds will improve with time will require red noise models for those pulsars.   Clearly, even with a 20-year observing span, it is unlikely that GWM detection will be made.

\begin{figure}
\centerline{\includegraphics[width=6cm,angle=-90]{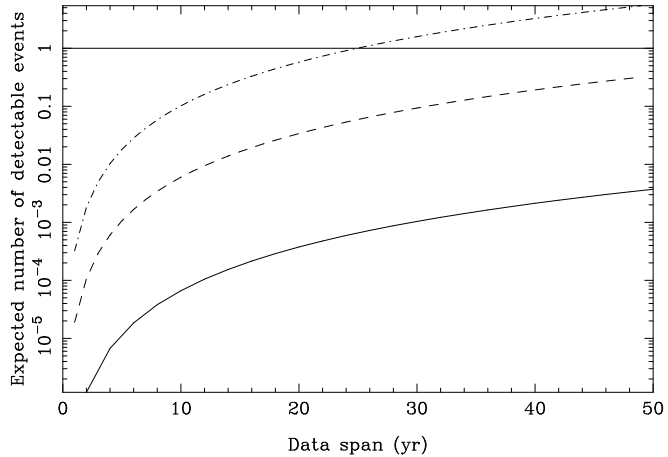}}
\caption{
\label{fg:predictDetect}
The expected number of detectable sources versus data span for idealised data sets containing: (solid line) 20 pulsars with 1$\mu$s rms timing residuals sampled every 20 days, (dashed line) the same, but for 100\,ns rms timing residuals and (dot-dashed line) 50 pulsars are observed with 50\,ns white noise and sampled once per week.}
\end{figure}

Secondly, we determine the parameters for idealized data sets. These idealised data sets have equal white noise for each pulsar, regular sampling and no red noise.  From the properties of such data sets it is possible to estimate the sensitivity of a given data set to GWM.  Combining equations 33 and 34 in van Haasteren \& Levin (2010) allows us to obtain an estimate of the detectable GWM amplitude
\begin{equation}
h^{\rm mem}_{\rm detectable} \sim 95 \sigma \sqrt{\frac{\Delta t}{N_p T_{\rm span}^3}}
\end{equation}
where $\sigma$ is the rms residual for the white data sets, $N_p$ the number of pulsars and $T_{\rm span}$ the data span.   Combining this with Equations~\ref{eq:pshirkov} and \ref{rate} allows us to estimate the number of  detectable sources expected for a given data set.  In Figure~\ref{fg:predictDetect} we show the expected number of sources as a function of data span assuming a GWB amplitude of $10^{-15}$ coming from a black hole population with $\mu = 10^{8}$M$_\odot$ at $z = 2$.  We first plot the expectation for a data set consisting of 20 pulsars with an rms timing residuals of 1$\mu$s and an observing cadence of 21 days (solid line).  This is similar to our actual current PPTA data set and highlights that extremely long data spans will be needed before we would expect make a GWM detection. We then try 20 pulsars with 100\,ns white noise and the same observing cadence (the result is shown as the dashed-line in the Figure).  It may be possible to obtain such a data set through the IPTA.  Again, long data sets (which are not significantly affected by red noise) will be needed before a detection is expected.  We finally plot (dot-dashed line) the expectation for 50 pulsars being observed at the 50\,ns level with weekly sampling.  This data set could be achievable on future telescopes such as FAST or the SKA.  In all cases, long data sets are required and we note that it is likely that the GW background will be detected well before bursts with memory are found.

These conclusions suggest that it is unlikely that GWM will be detected in the near future. However, it is still important that bursts with memory are searched for as our data sets get longer because:
\begin{itemize}
\item As described in this paper, searching for GWM events is reasonably straightforward and the techniques are built into our existing software packages.
\item The astrophysical predictions may be incorrect.
\item GWM events may occur from sources other than black hole binary systems (for instance, from cosmic strings) and those sources may have a higher event rate.
\item As shown in this paper, searching for GWM events leads to an improved understanding of the noise processes within a given data set.
\end{itemize}

Throughout this paper we have considered GWM events passing through the solar neighbourhood.  A GWM event passing a pulsar will induce a glitch event in that pulsar alone.  It would be extremely difficult to prove that any such event in a single pulsar was caused by a GWM, but searching for such events can be used to improve the bounds on the event rates.   Whereas in this paper we can only constrain events within our data span of $\sim 5$\,yr, by searching for GWM events in each pulsar independently we would be able to place a constraint over a time scale of $\sim 20 \times 5 = 100$\,yr. Such searches have been described by Cordes \& Jenet 2012 and Madison et al. 2014.

\section{Conclusions}

We have presented straightforward algorithms for detecting the presence of a GWM burst and for bounding the event rate of GWM bursts when no burst is detected.  No GWM burst events were detected when applying the algorithm to the first PPTA data release.

The weakest GWM burst we could have detected has an amplitude $\sim 2 \times 10^{-14}$. Such a burst would have to occur in a small region of the sky to be detected. By comparison, a burst strong enough that we would have detected it anywhere in the sky would have an amplitude greater than $2.4 \times 10^{-13}$. The event rate for such bursts must be less than 0.75 per year.  These bounds do not significantly constrain models for supermassive black hole binary coalescence rates, and, although the bounds will improve with time, the  detection of a GWM event may take many years.

\section{Acknowledgements}

The Parkes radio telescope is part of the Australia Telescope, which is funded by the Commonwealth of Australia for operation as a National Facility managed by the Commonwealth Scientific and Industrial Research Organisation (CSIRO). This work is supported by National Basic Research Program of China (973 Program 2012CB821800), National Natural Science Foundation of
China (Nos. 11403086, U1431107, 11173041, 11173042, 11373006, and 11203063), and West Light Foundation of CAS (No.XBBS201322, No.XBBS201223, and XBBS201123). GH, LW and YL acknowledge support from the Australian Research Council (ARC) Future Fellowship. V.R. is a recipient of a John Stocker Postgraduate Scholarship from the Science and Industry Endowment Fund. D.R.M. received support from the NSF/PIRE Grant 0968296. X-JZ acknowledges the support of an University Postgraduate Award at UWA.

\bibliographystyle{mn2e}

\appendix

\section{The \textsc{efacEquad} plugin}

The \textsc{efacEquad} plugin is used to produce values of \textsc{EFACs} and \textsc{EQUADs} for a given data set.  It relies on the user having identified specific backend and receiver combinations that are expected to have the same  \textsc{EFAC} and \textsc{EQUAD} values.  Different combinations are known as different ``groups" and are uniquely identified using \textsc{tempo2} flags.  The plugin accepts various command line arguments. The most common usage is:
\begin{verbatim}
tempo2 -gr efacEquad -f psr.par psr.tim
   -flag <flagID> -plot
\end{verbatim}
where \verb|flagID| is the flag identifying each group.

The plugin then:
\begin{itemize}
\item Switches off all fitting within the parameter file before turning on fits for the pulse frequency (F0) and its first time derivative (F1).
\item The red noise is modelled using a constrained, linear interpolation method (known within \textsc{tempo2} as \textsc{IFUNCS}).   By default, the linear interpolation is based on a 100\,d grid.
\item The data are refitted using the linear interpolation, F0 and F1.
\item The post-fit residuals are extracted for each group in turn.
\item For each group the reduced-$\chi^2$ of the fit is determined. If the reduced-$\chi^2 < 1$ then the uncertainties for that group are decreased by $\sqrt{\chi^2_r}$, otherwise the following process is carried out.
\begin{itemize}
\item The distribution of normalised residuals (i.e., the residual divided by its uncertainty) is determined and compared (currently using a Kolmogorov-Smirnov test) with a Gaussian distribution with zero mean and unit variance.
\item The normalised residuals are re-calculated by modifying the ToA uncertainties with specific \textsc{EFAC} and \textsc{EQUAD} values.  The values are chosen from a grid.  For each grid point the Kolmogorov-Smirnov test probability is recorded and graphically plotted.
\item The grid point (\textsc{EFAC},\textsc{EQUAD}) that leads to the best match according to the Kolmogorov-Smirnov test is output and used in subsequent processing.
\end{itemize}
\end{itemize}

\section{The EFAC and EQUAD values}

In order to calculate EFAC and EQUAD values the observations were first grouped into backend/receiver combinations that we expect to have the same properties.  These are listed in Table~\ref{tb:groups}.

\begin{table}
\caption{The groupings used when calculating EFAC and EQUAD values}\label{tb:groups}
\begin{center}\begin{tabular}{lll}\hline
Receiver & Backend & Group label \\ \hline
MULTI & cpsr2n & 3 \\
MULTI & WBCORR & 4 \\
10CM &WBCORR & 6  \\
MULTI & CPSR2m  & 7  \\
MULTI & CPSR2n & 7 \\
10CM & PDFB1 & 8 \\
MULTI & PDFB1 & 9 \\
H-OH & PDFB1 & 9 \\
MULTI & PDFB1  & 9d1 \\
MULTI & PDFB2  & 9d2 \\
MULTI & PDFB2  & 9\_d2 \\
MULTI & PDFB3  & 9  \\
MULTI & PDFB2  & 9 \\
MULTI &PDFB3  & 9 \\
MULTI & PDFB4  & 9 \\
10CM & PDFB4 &  10 \\
10CM & PDFB2 & 10 \\
10CM & PDFB2 &  10a \\
MULTI & APSR & 13 \\
10CM& PDFB4  & dfbs \\
10CM& PDFB1  & dfb1 \\ \hline
\end{tabular}\end{center}
\end{table}

The values for each pulsar are as follows:

\begin{verbatim}
PSR J0437-4715
T2EQUAD -group 8 0.033
T2EQUAD -group 10a 0.047
T2EFAC -group 10 1.7
T2EQUAD -group 10 0.008
EQUAD 0.080

PSR J0613-0200
T2EQUAD -group 7 0.6
T2EFAC -group 9 1.2
T2EQUAD -group 9 0.2

PSR J0711-6830
T2EFAC -group 4 0.667076
T2EFAC -group 7 0.94107
T2EQUAD -group 9 1

PSR J1022+1001
T2EFAC -group 7 1.3
T2EQUAD -group 7 0.6
T2EQUAD -group 9d1 1.4
T2EFAC -group 9d2 1.7
T2EQUAD -group 9d2 0.5
T2EFAC -group 9 1.9
T2EQUAD -group 9 0.4
EQUAD 0.5

PSR J1024-0719
T2EFAC -group 4 0.956148
T2EFAC -group 7 1.1
T2EQUAD -group 9 0.3

PSR J1045-4509
T2EQUAD -group 4 0.9
T2EFAC -group 7 0.96299
T2EFAC -group 4 1.2

PSR J1600-3053
T2EQUAD -group 7 0.4
T2EFAC -group 4 0.745517
T2EFAC -group 9 1.1
T2EQUAD -group 9 0.3

PSR J1603-7202
T2EQUAD -group 7 1.2
T2EQUAD -group 9 0.7
T2EQUAD -group 9_d2 4
T2EFAC -group 9 1.1

PSR J1643-1224
T2EFAC -group dfb1 0.922936
T2EFAC -group dfbs 0.985818

PSR J1713+0747
T2EQUAD -group 6 0.5
T2EFAC -group 8 1.4
T2EFAC -group 10 1.6

PSR J1730-2304
T2EFAC -group 7 1.2
T2EQUAD -group 9 1

J1732-5049
T2EQUAD -group 9 0.8
T2EFAC -group 7 1.3
T2EQUAD -group 7 0.3

PSR J1744-1134
T2EFAC -group 7 1.3
T2EQUAD -group 7 0.1
T2EFAC -group 4 2.6
T2EFAC -group 9 1.1
T2EQUAD -group 9 0.2

PSR J1857+0943
T2EFAC -group 4 0.741308
T2EFAC -group 7 0.891042
T2EQUAD -group 9 0.1

PSR J1909-3744
EQUAD  0.1

PSR J1939-3744
T2EFAC -group 7 1.3
T2EFAC -group 9 2.3
T2EQUAD -group 9 0.1
T2EFAC -group 13 2.4
T2EQUAD -group 13 0.1

PSR J2124-3358
T2EFAC -group 4 1.1
T2EQUAD -group 4 2.8
T2EFAC -group 3 0.707916
T2EQUAD -group 9 0.6
T2EFAC -group 13 2.6

PSR J2129-5721
T2EFAC -group 7 0.897594
T2EFAC -group 9 0.904696

PSR J2145-0750
T2EQUAD -group 4 0.8
T2EQUAD -group 7 0.5
T2EQUAD -group 9 0.6
EQUAD 0.800000

\end{verbatim}

\section{Tempo2 and GWM}

As part of this work, various updates have been made to the \textsc{tempo2} software packages. This updates are available in the current distribution (\url{http://sourceforge.net/projects/tempo2/}).

A GWM event can be defined in a pulsar timing model using:
\begin{verbatim}
GWM_AMP <amp> <fitflag>
GWM_POSITION <ra> <dec>
GWM_EPOCH <mjd>
GWM_PHI <zeta>
\end{verbatim}
where \verb|fitflag| is usually set to `2' to indicate a global fit.  The position ($\alpha_g,\delta_g$) and the position angle (in this paper described as $\zeta$) are given in radians.  Instead of using \verb|GWM_AMP| and \verb|GWM_PHI| we have found that the following parameterisation is more useful:
\begin{verbatim}
GWM_A1 <amp1> <fitflag>
GWM_A2 <amp2> <fitflag>
GWM_POSITION <ra> <dec>
GWM_EPOCH <mjd>
\end{verbatim}

These parameters can be included when fitting or for simulating data sets.   In this paper all fits were carried out using the Cholesky routines using a command similar to:
\begin{verbatim}
tempo2 -f psr1.par psr1.tim -f psr2.par psr2.tim ...
  -global global.par -dcf model.dat
\end{verbatim}
where the red noise model file was:
\begin{verbatim}
MODEL T2
PSR J0437-4715
 MODEL T2PowerLaw 3 4.34384e-29 0.2
PSR J0613-0200
 MODEL T2PowerLaw 5 5.52754e-28 0.4
PSR J1022+1001
 MODEL T2PowerLaw 3 1.78292e-27 0.5
PSR J1024-0719
 MODEL T2PowerLaw 4 3.96149e-27 0.2
PSR J1045-4509
 MODEL T2PowerLaw 2.5 2.0376e-26 0.2
PSR J1600-3053
 MODEL T2PowerLaw 4 3.81388e-27 0.2
PSR J1603-7202
 MODEL T2PowerLaw 3 2.27845e-27 0.2
PSR J1643-1224
 MODEL T2PowerLaw 4 1.5e-26 0.2
PSR J1713+0747
 MODEL T2PowerLaw 4 4.13154e-28 0.2
PSR J1824-2452A
 MODEL T2PowerLaw 4 1.30232e-25 0.2
PSR J1939+2134
 MODEL T2PowerLaw 2.5 2.86815e-27 0.2
PSR J2129-5721
 MODEL T2PowerLaw 4 3.56769e-27 0.2
 \end{verbatim}

\end{document}